\documentclass[12pt,prl,superscriptaddress,showpacs,preprintnumbers,
amsmath,amssymb]{revtex4}

\usepackage{graphicx} 
\usepackage{dcolumn} 

\def\bk{{\bf k}}

\def\bq{{\bf q}}
\def\br{{\bf r}}

\def\b0{{\bf 0}}

\def\bra{\langle}
\def\ket{\rangle}

\def\eps{\epsilon}

\def\Lam{\Lambda}
\def\om{\omega}

\def\sg{\sigma}

\begin{document}

\title{Nematic Quantum Criticality Without Order}

\author{H.~Yamase}
\affiliation{National Institute for Materials Science, 
 Tsukuba 305-0047, Japan}
\author{P.~Jakubczyk}
\affiliation{Institute of Theoretical Physics, Faculty of Physics, 
 University of Warsaw, Ho\.za 69, 00-681 Warsaw, Poland}
\author{W.~Metzner}
\affiliation{Max-Planck-Institute for Solid State Research,
 D-70569 Stuttgart, Germany}

\date{\today}

\begin{abstract}
We consider a two-dimensional interacting Fermi system which displays a
nematic phase within mean-field theory. The system is analyzed using a 
non-perturbative renormalization-group scheme. 
We find that order-parameter fluctuations can suppress the nematic order 
obtained in mean-field theory even at zero temperature.
For a suitable choice of parameters a quantum critical point surrounded
by a disordered phase can be realized, giving rise to quantum critical
behavior in the absence of an ordered regime in the phase diagram.
\end{abstract}

\maketitle


Classical phase transitions are controlled by an energy-entropy balance,
and the critical behavior of continuous transitions is caused by thermal
fluctuations.
By contrast, quantum phase transitions occur at zero temperature and 
are driven by quantum effects \cite{sachdev99}.
In the last decade there has been growing interest in quantum phase
transitions in metallic systems, and especially in deviations from
conventional Fermi liquid behavior due to quantum critical fluctuations
near a quantum critical point \cite{vojta03,loehneysen07}.

A widely used approach to approximately determine equilibrium phase 
diagrams of interacting systems is the mean-field theory (MFT). 
In numerous cases this approach yields a qualitatively correct description 
of the phase diagram in question. 
There are however also results revealing the inadequacy of this approximation
in certain situations, in particular in low-dimensional systems. 
The Mermin-Wagner theorem \cite{mermin66} delivers a well-known, transparent 
and rigorous example of such a result for finite temperatures and spatial 
dimension $d \leq 2$. 
A recent example for the case $T=0$ was provided by Greenblatt {\it et al.}\
\cite{greenblatt09}, establishing rigorously the Imry-Ma rounding 
phenomenon due to disorder for quantum phase transitions. 
Many other (mostly approximate) results point toward the very important role 
played by quantum fluctuations in $d=2$ and $T \geq 0$
\cite{sachdev99,vojta03,loehneysen07}. 
Also for a number of systems in $d=3$ and $T \geq 0$ (see e.g.\ Ref.\ 
\cite{belitz05}) fluctuations are known to severely influence the phase 
diagram, for example by turning a continuous transition into a first-order 
one, or vice-versa~\cite{fucito81}.

There is considerable interest in two-dimensional itinerant 
electron systems with interactions favoring a $d$-wave Pomeranchuk 
instability \cite{yamase00,halboth00} leading to a so-called 
electronic nematic state \cite{fradkin10}, where the point group 
symmetry of the underlying square lattice is spontaneously broken.
There is growing evidence that a nematic phase is realized in the 
ruthenate compound $\rm Sr_3Ru_2O_7$ \cite{ruthenate}.
Signatures of a nematic instablility have also been observed in cuprates
\cite{cuprates} and other strongly correlated electron materials.

The nematic transition was found to be generically of first order at 
zero and sufficiently low temperatures for a number of models solved 
within the mean-field approximation \cite{kee03,khavkine04,yamase05}, 
while the transition to the normal metallic state becomes continuous 
only for higher temperatures, above a tricritical temperature.
In a phase diagram spanned by the chemical potential and temperature, 
the nematic phase resides within a dome-shaped region around van Hove
filling.
Due to the vicinity of the van Hove singularity in the density of
states, at zero temperature the Landau expansion of the mean-field 
potential in powers of the order parameter $\phi$ exhibits negative 
interactions at any finite order in $\phi$. 
Hence, fluctuation effects cannot be computed perturbatively.
This problem was recently overcome by using non-perturbative flow
equations derived within the functional renormalization group (RG) 
framework, and it was shown that order-parameter fluctuations may 
actually suppress the first order character of the transition even 
at $T=0$, such that a continuous quantum phase transition accompanied 
by quantum critical fluctuations emerges \cite{jakubczyk09}. 

In this letter we show that the impact of quantum fluctuations can be 
even more drastic. 
We show that the nematic order present within MFT can be completely 
swept out by the order-parameter fluctuations. 
We also observe that one can tune the system to a scenario where
a singular quantum critical point is present, while no ordered phase
occurs whatsoever. 

We analyse a system of itinerant electrons on a square lattice with
a Hamiltonian of the form \cite{metzner03}
\begin{equation}
 H = \sum_{\bk} \eps_{\bk} n_{\bk} +
 \frac{1}{2L} \sum_{\bk,\bk',\bq} f_{\bk\bk'}(\bq) \,
 n_{\bk}(\bq) \, n_{\bk'}(-\bq) \; ,
\label{f-model}
\end{equation}
featuring the $d$-wave forward scattering interaction
\begin{equation}
 f_{\bk\bk'}(\bq) = - g(\bq) d_{\bk} d_{\bk'} \; ,
\end{equation}
where $d_{\bk} = \cos k_x - \cos k_y$ is a form factor with
$d_{x^2-y^2}$ symmetry. The coupling function $g(\bq) \geq 0$
has a maximum at $\bq=\b0$ and is restricted to small momentum 
transfers by a momentum cutoff. 
We introduced  $n_{\bk}(\bq) = \sum_{\sg} 
 c^{\dag}_{\bk-\bq/2,\sg} c^{\phantom\dag}_{\bk+\bq/2,\sg} \,$, 
where $c^{\dag}_{\bk\sg}$ ($c_{\bk\sg}$) creates (annihilates)
an electron with momentum $\bk$ and spin orientation $\sg$,
and $L$ denotes the number of lattice sites.
Note that $n_{\bk} = n_{\bk}(\b0)$.
We consider a dispersion given by
\begin{equation}
 \eps_{\bk} = - 2t(\cos k_x + \cos k_y) - 4t' \cos k_x \cos k_y 
 - 2t''(\cos 2 k_x + \cos 2 k_y) \; ,
\end{equation}
with nearest, next-nearest, and third-nearest hopping amplitudes
$t$, $t'$, and $t''$, respectively. 

For sufficiently large $g = g(\b0)$ the interaction drives a 
$d$-wave Pomeranchuk instability leading to a nematic state with 
broken orientation symmetry, which can be described by the order 
parameter
\begin{equation}
 \phi = \frac{g}{L} \sum_{\bk} d_{\bk} \bra n_{\bk} \ket \; .
\end{equation}
Introducing a fluctuating order parameter field via a 
Hubbard-Stratonovich transformation, the fermionic interaction 
term in Eq.~(\ref{f-model}) is decoupled, and the fermionic 
degrees of freedom can be integrated out. 
Retaining only the leading momentum and frequency dependencies of the
two-point function and neglecting such dependencies in the higher-order
vertex functions \cite{hertz76}, one obtains the following action:
 \begin{eqnarray}
 {\cal S}[\phi] = 
 \frac{T}{2} \sum_{\om_n} \int \frac{d^2q}{(2\pi)^2} \, 
 \phi_{\bq,\om_n} \left( A_0 \frac{|\omega_{n}|}{|\bq|}
 + Z_0 \bq^{2} \right) \phi_{-\bq,-\om_n} + {\cal U}[\phi] \; ,
 \label{action}
\end{eqnarray}
where $\phi_{\bq,\om_n}$ is the momentum representation of the 
order parameter field $\phi$, and $\omega_n = 2\pi n T$ with integer 
$n$ denotes the (bosonic) Matsubara frequencies.
The momenta and frequencies contributing to the action ${\cal S}[\phi]$ 
are restricted by an ultraviolet cutoff $\Lambda_0$ to the
region $A_0 \frac{|\omega_{n}|}{|\bq|} + Z_0 \bq^{2} 
\leq Z_0 \Lambda_0^2$. 
In the fermionic representation, $\Lambda_0$ is related to the maximal 
momentum transfer allowed by the interaction in Eq.~(\ref{f-model}).
The values of the prefactors $A_0$ and $Z_0$ are related to the 
$d$-wave particle-hole bubble in presence of a (mean-field) order
parameter $\phi_0$, and $Z_0$ depends also on the specific form of the 
function $g(\bq)$, as discussed in Refs.~\cite{jakubczyk09,dellanna06}.
We do not treat a possible dependence of $A_0$ and $Z_0$ on the 
direction of $\bq$, such that the values we insert should be viewed 
as an angular average.
The effective potential ${\cal U}[\phi]$ is given by 
\begin{equation}
 {\cal U}[\phi] = 
 \int_0^{\frac{1}{T}} \! d\tau \int d^2 r \, U(\phi(\br,\tau)) \; ,
\label{potential}
\end{equation}
where 
\begin{equation}
 U(\phi) = 
 \frac{\phi^2}{2g} - 2T \int \frac{d^2 k}{(2\pi)^2} \,
 \ln\left( 1 + e^{-(\eps_{\bk} - \phi d_{\bk} - \mu)/T}
 \right) \; .
\label{potential_i}
\end{equation}
An action of the form (\ref{action}) was already considered in 
Ref.~\cite{jakubczyk09}. 
A polynomial expansion of $U(\phi)$ yields negative coefficients at any 
order in $\phi$ and is therefore inadequate here.  
Our calculations are carried out within the one-particle irreducible 
version of the functional RG \cite{wetterich93}. 
The analysis amounts to solving the flow equations for $U(\phi)$ and
$Z$ derived in Ref.~\cite{jakubczyk09}. 
Here we only remark that we employ a truncation of the functional RG 
based on the approximation that the effective action retains the local 
structure given by Eq.~(\ref{potential}). 
We do not impose any ansatz for $U(\phi)$ and therefore keep track 
of the renormalization of infinitely many couplings. The flow of the 
coefficient $Z$ captures the anomalous dimension of the order parameter 
field at criticality. The truncation level is analogous to the next-to-leading 
order in the derivative expansion, which was developed in the context of 
classical criticality \cite{berges02,delamotte04}.

We now present our results, which are obtained from the solution of the RG 
flow equations for the following set of parameters: 
$t=1$, $t'=-1/6$, $t''=1/5$, $g=1/2$, $A_0 = 1$, $Z_0 = 10$, and for 
different values of $\Lambda_0$. 
For this choice of hoppings the density of states is strongly asymmetric
and the chemical potential corresponding to the maximal $T_c$ is 
separated from the van Hove energy ($\mu_{\rm vH} = -1.444$).
The relatively large value of $Z_0$ corresponds to a pronounced peak
of $g(\bq)$ at $\bq = 0$, which favors the nematic instability 
compared to instabilities at a finite $\bq$.
Fluctuations reduce the magnitude of the order parameter compared to
the mean-field value, as expected, and the nematic region in the phase
diagram shrinks (see Fig.~1). The suppression is naturally stronger for
larger cutoffs $\Lam_0$.
\begin{figure}[ht]
\centerline{\includegraphics[width = 10cm]{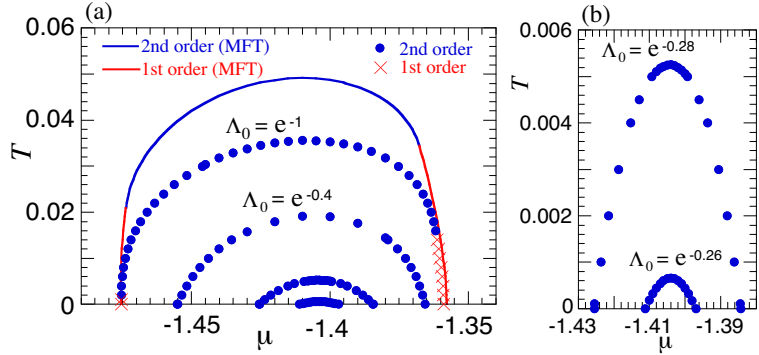}}
\caption{(color online). Critical temperature for the nematic transition
 as a function of the chemical potential $\mu$. 
 Left panel (a) Critical temperature in mean-field theory (MFT) and 
 in the presence of fluctuations with four different choices of 
 the ultraviolet (momentum transfer) cutoff, 
 $\Lam_0 = e^{-1}, e^{-0.4}, e^{-0.28}, e^{-0.26}$.
 Right panel (b) Enlargement of the results for 
 $\Lam_0 = e^{-0.28}, e^{-0.26}$.}
\label{fig1}
\end{figure}
For sufficiently large $\Lambda_0$ the nematic phase is located to the 
right of the van Hove filling. 
When $\Lambda_0$ exceeds a critical value $\Lambda_0^*$, no ordering 
occurs whatsoever. 
By continuity, for $\Lambda_0 = \Lambda_0^*$ the phase diagram exhibits 
an isolated quantum critical point at $\mu = \mu_{\rm qc}^*$ and 
$T=0$, while the system is disordered in the entire $(\mu,T)$ plane. 
In the vicinity of the point $(\mu=\mu_{\rm qc}^*,T=0)$ 
the system displays soft fluctuations characterized by a large 
correlation length. For the present choice of parameters we find 
$\mu_{\rm qc}^* \approx -1.404$ and $\Lambda_0^* \approx e^{-0.2574}$.
Note that for smaller values of $g$ the order would be destroyed 
already at smaller $\Lam_0$.

We have also computed the behavior of the correlation length $\xi$ upon 
approaching the quantum critical points at the edges of the nematic
dome for $\Lam_0 < \Lam_0^*$, and for the isolated quantum critical
point in the special case $\Lam_0 = \Lam_0^*$.
The correlation length in the symmetric phase is determined by the 
curvature of the potential $U''(0)$ and the renormalization constant 
$Z$ at the end of the flow via $\xi^2 = Z/U''(0)$. 
For $\Lam_0 < \Lam_0^*$ the results are consistent with the predictions 
of Hertz-Millis theory \cite{millis93} for a two-dimensional
system with a dynamical exponent $z=3$, namely 
$\xi^2 \sim |T\log T|^{-1}$ at fixed $\mu = \mu_{\rm qc}$ 
and $\xi^2 \sim |\mu-\mu_{\rm qc}|^{-1}$ at $T=0$.  

The scenario occurring for $\Lambda_0 = \Lambda_0^*$ requires 
fine-tuning. However, features resembling quantum-critical behavior 
persist also for $\Lambda_0 > \Lambda_0^*$. 
In Fig.~2 we plot the correlation length $\xi$ as a function of $\mu$ 
at $T=0$ for a sequence of values of $\Lambda_0$ approaching 
$\Lambda_0^*$ from above.
\begin{figure}[ht]
\centerline{\includegraphics[width = 7cm]{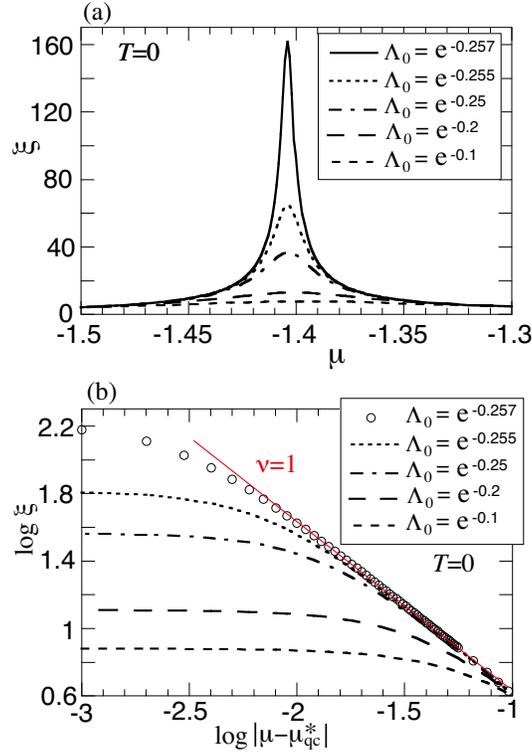}}
\caption{(color online) Top (a) Correlation length $\xi$ as a 
 function of $\mu$ at $T=0$ for various values of $\Lam_0 > \Lam_0^*$. 
 Bottom (b) Double logarithmic plot of $\xi$ versus 
 $\mu - \mu_{\rm qc}^*$ and fit to the exponent $\nu = 1$.}
\label{fig2}
\end{figure}
The peak formed around $\mu = \mu_{\rm qc}^*$ becomes rapidly 
pronounced as $\Lambda_0 \to \Lambda_0^*$. 
Remarkably, although the quantum critical point occurring for 
$\Lambda_0 = \Lambda_0^*$ is Gaussian, the exponent $\nu$ describing 
the divergence of $\xi$ at $\mu \to \mu_{\rm qc}^*$ takes the value 
$1$ instead of the usual value $\frac{1}{2}$. 
This is easily understood by noticing that for $\Lambda_0 = \Lambda_0^*$ 
the quadratic mass term in the Landau expansion must not take negative 
values in the absence of an ordered phase and therefore cannot be 
proportional to $\mu - \mu_{\rm qc}^*$ as in the usual cases.
The natural $\mu$-dependence of the quadratic term is therefore 
quadratic in $\mu - \mu_{\rm qc}^*$ near $\mu_{\rm qc}^*$, which
leads immediately to $\nu = 1$.
This peculiarity does not influence the behavior when the singularity
is cut off by temperature. 
In fact for fixed $\mu = \mu_{\rm qc}^*$ we recover the usual 
\cite{millis93} behavior $\xi \sim |T\log T|^{-1/2}$, as shown in
Fig.~3.
\begin{figure}[ht]
\centerline{\includegraphics[width = 7cm]{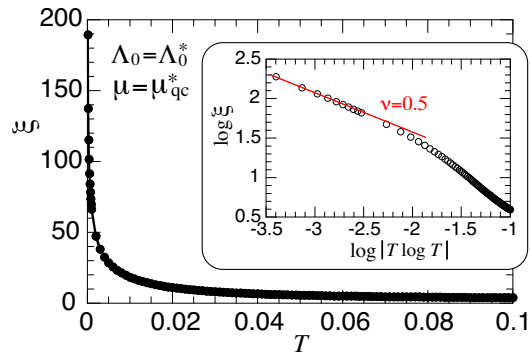}}
\caption{(color online) Temperature dependence of the correlation length
 $\xi$ for the special case $\mu = \mu_{\rm qc}^*$ and 
 $\Lam_0 = \Lam_0^*$. Inset: Double logarithmic plot exhibiting the
 expected behavior $\xi \sim |T\log T|^{-1/2}$ for small $T$.}
\label{fig3}
\end{figure}
For $\Lambda_0 > \Lambda_0^*$ no singular point in the phase 
diagram occurs, but the correlation length takes sizable values in the 
vicinity of the point $(\mu=\mu_{\rm qc}^*,T=0)$ as long as $\Lambda_0$
is not significantly larger than $\Lambda_0^*$.

One may be surprised to find the nematic order fully eliminated 
by fluctuations even at van Hove filling. 
In MFT, the diverging density of states at van Hove filling leads 
to an instability of the symmetric phase at any (arbitrarily small) 
coupling $g$ \cite{yamase05}. 
One would expect this to remain true also beyond MFT, provided the 
system remains a Fermi liquid.
Inversely, the absence of nematic order at van Hove filling despite
the presence of a $d$-wave attraction $g$ would imply the absence
of fermionic quasi-particles, that is, non-Fermi liquid behavior.
Unlike the situation at the quantum critical point, that non-Fermi 
liquid behavior is not related to critical fluctuations.
Working with an effective order parameter action we do not have
direct access to fermionic properties.
However, the breakdown of Fermi liquid theory at van Hove 
points was already diagnosed many years ago by Dzyaloshinskii 
\cite{dzyaloshinskii96}. In the light of that work the complete
suppression of order by fluctuations even at van Hove filling
is a consistent possibility.

In summary, we have analyzed a two-dimensional interacting electron
system which exhibits a nematic phase in mean-field theory.
We have shown that fluctuations can extinguish the nematic order
completely from the phase diagram spanned by the chemical potential
and temperature. 
For a special choice of parameters a quantum critical point can be
realized in the absence of order.
Despite being realized as a consequence of strong quantum 
fluctuations, the quantum critical point itself is Gaussian.
The temperature dependence of the correlation length in the quantum
critical regime above this isolated quantum critical point follows
the conventional behavior known for Gaussian quantum criticality.
However, at zero temperature the correlation length diverges with
an exponent $\nu = 1$ instead of $\nu = \frac{1}{2}$ upon approaching 
the critical chemical potential.
While the present calculation was carried out for a specific 
interaction triggering a nematic phase, an analysis of fluctuation 
effects in other two-dimensional systems described by an action of 
the form (\ref{action}) can be performed along the same lines. 

\vspace*{1cm}

\begin{acknowledgments}
We are grateful to J.\ Bauer, K.\ Byczuk, T.\ Holder, and 
C.\ Husemann for valuable discussions.
This work was supported by the German Research Foundation through
the research group FOR 723.
H.Y. was also supported by a Grant-in-Aid for Scientific Research 
from Monkasho.  
\end{acknowledgments}


\end{document}